\documentstyle[aps,prl,psfig,twocolumn]{revtex}
\begin{document}
\def\be{\begin{equation}}
\def\ee{\end{equation}}
\def\bc{\begin{center}}
\def\ec{\end{center}}
\def\bea{\begin{eqnarray}}
\def\eea{\end{eqnarray}}
\draft

\title{High dimensional behavior of the Kardar-Parisi-Zhang 
growth dynamics}

\author{C. Castellano$^{(1)}$,  A. Gabrielli$^{(2,3)}$, M. Marsili$^{(4)}$,
M. A.  Mu{\~{n}}oz$^{(2)}$, 
and L. Pietronero$^{(1,2)}$}
\address{$^{(1)}$ The Abdus Salam International 
Center for Theoretical Physics, P. O. Box 586,
I-34100 Trieste, Italy\\
$^{(2)}$ Dipartimento di Fisica and Unit\`a INFM,
Universit\`a di Roma ``La Sapienza'', I-00185 Roma, Italy\\
 $^{(3)}$   Dipartimento di Fisica Universit\`a di Roma ``Tor Vergata'',
I-00133 Roma, Italy\\
$^{(4)}$
International School for Advanced Studies
(SISSA) and Unit\`a INFM, via Beirut 2-4, Trieste I-34014, Italy
}

\maketitle 

\begin{abstract}
We investigate analytically the large dimensional behavior 
of the Kardar-Parisi-Zhang (KPZ) dynamics of surface growth
using a recently proposed non-perturbative renormalization 
for self-affine surface dynamics.
Within this framework, we show that the roughness exponent 
$\alpha$ decays not faster than $\alpha\sim 1/d$ for large $d$.
This implies the absence of a finite upper critical dimension.
\end{abstract}
\pacs{PACS numbers: 05.40.+j,64.60.Ak,05.70.Ln,68.35.Fx}
\narrowtext
      
The study of the non equilibrium dynamics of rough surfaces and
interfaces has received a great deal of attention in the last 
years\cite{HZ,Laszlo}. 
Both theoretically and experimentally many efforts have been
devoted to single out the traits and features shared by apparently
different phenomena. In this context, by analogy with equilibrium 
statistical mechanics, the search for universality 
classes is a central task. The Kardar-Parisi-Zhang \cite{kpz} 
equation (KPZ) is, for surface growth, the main contribution in 
this direction. It is the minimal Langevin equation
capturing the essence of many different growth models beyond 
the Gaussian linear theory \cite{HZ,Laszlo}. It reads 
\be
{\partial h(x,t) \over \partial t} = \nu \nabla^2 h + {\lambda \over 2}
(\nabla h)^2 + \eta(x,t).
\ee
where $h(x,t)$ is the surface profile, $x$ is the position in a
$d$-dimensional substrate, $\eta$ is a Gaussian white noise,
$\nu$ and $\lambda$ are constants.
The KPZ equation also describes the behavior of directed polymers 
in random media \cite{HZ}, systems with multiplicative noise \cite{MN},
and it is related to the Burgers equation \cite{burger}. 

A central quantity of interest is the roughness $W(L)$ of
a system of linear size $L$, defined as 
\be 
 W^2(L)=\frac{1}{L^d}\sum_{x} [h(x,t) -\bar{h}]^2.
\label{roughness}
\ee 
where $ \bar{h} = (1/L^d)\sum_{x} h(x,t)$.
In many seemingly unrelated growth processes the large scale 
properties of the roughness are observed to be scale 
invariant and universal; i.e. in the stationary state 
$W(L) \sim L^\alpha$ and correlations decay on a 
typical time $t_s\sim L^z$, with universal exponents 
$\alpha$ and $z$. These critical exponents are not 
independent, as a consequence
of the Galilean invariance of the related Burgers equation
\cite{HZ,burger} $\alpha + z=2$. It is thus sufficient to 
focus the attention on one exponent, say $\alpha$.

The theoretical analysis of the KPZ is extremely difficult.
Apart from the $d=1$ case, where a special symmetry makes
an exact solution possible with $\alpha=1/2$, the situation 
is still quite controversial 
despite the large effort devoted to the problem.
In particular, the fundamental issue of the existence of
an upper critical dimension $d_c$, above which the exponents 
recover their  mean-field (or infinite dimensional) values
($\alpha=0$)\cite{infinited}, is 
highly debated\cite{debate}. The application 
of field theoretical tools presents an inherent 
problem: one indeed finds that the fixed point controlling the 
rough phase of the KPZ is not accessible to perturbation 
expansion in $\lambda$;
this fact renders standard field theoretical tools
inadequate for this problem. Early applications of 
non-perturbative methods such as functional renormalization
group\cite{hh} and Flory-type arguments \cite{Flory} suggested
that $d_c=4$, in agreement with a $1/d$-expansion\cite{1/d} 
around the $d=\infty$ limit. Later the mode-coupling approximation 
led to contradictory results suggesting the existence of a finite
$d_c$ \cite{MCDC} or $d_c=\infty$ \cite{Yuhai}. Arguments
for a finite $d_c$ based on directed\cite{DP} or
invasion\cite{CMB} percolation have also been proposed.
More recently a detailed analysis of a $d=2+\epsilon$
perturbative expansion revealed a singularity at
$d=4$\cite{Nuclear}, leading L\"assig to the 
conclusion that $d_c=4$ is the upper critical dimension of 
the KPZ dynamics \cite{Lassig}.

Numerical simulations of models in the KPZ universality 
class markedly disagree with this last conclusion \cite{debate},
showing that $\alpha>0$ at least up to $d=7$\cite{Ala}.
In particular numerical results suggest a large-$d$ behavior
$\alpha\sim 1/d$ in agreement with early conjectures 
\cite[p. 75]{Laszlo}. 
Both of these conclusions were confirmed by a recently
proposed renormalization group (RG) approach \cite{CML}.
The key idea of this approach is that the geometric scaling 
of the growing surface can be ascribed to a scale 
invariant dynamic process, which builds the same
correlations at all length-scales. This {\em scale
invariant dynamics} is the fixed point of the RG 
transformation, which is derived by consistency requirements 
of the description of the same system at two different 
scales. Analogous ideas, implemented {\em via} a real space RG,
have proved to be quite powerful to investigate the
critical properties of non-equilibrium, strong coupling
problems\cite{sdi}. The implicit nature of the RG transformation, 
which is similar in spirit to the idea of phenomenological 
RG\cite{PhRG}, allows us to avoid the use of hierarchical 
lattices, a source of incontrollable 
approximations, specially in high dimensions. 
Remarkably, the exponents predicted by the RG are in excellent 
agreement with numerical simulations up to $d=7$.

In this Letter we analyze the large-$d$ behavior of this
RG approach and  show that it predicts that the roughness 
exponent $\alpha$ vanishes not faster than $1/d$ for $d\gg 1$. 
This rules out the existence of a finite upper critical 
dimension.
In what follows we expose the essential concepts of this method
and apply it to the analytical study of the KPZ dynamics 
in the large-$d$ limit. 

Consider a growing surface, whose dynamics, at the microscopic
scale, is defined in terms of a stochastic equation, such as
Eq.~(1), or by a discrete model. If we partition the
$d+1$-dimensional space in cells of lateral size 
$L_k=2^k L_0$ and vertical size $h_k$, we obtain a 
{\em static} description of the surface at the coarse
grained scale $L_k$: With some majority rule each block is 
declared to be empty or filled. For each substrate cell $i$
the number $h(i)$ of filled blocks on top of it identifies the
interface configuration, in units $h_k$, at scale $L_k$.
Note that $h_k$ is an independent parameter of the 
static description. Scale invariance implies that if 
$h_k$ is properly chosen, the coarse grained system 
looks similar at all (large enough) length-scales $L_k$. 
The optimal geometric description, which best exhibits scale 
invariance, in our case, is that with $h_k\propto W(L_k)\sim 
L_k^\alpha$ of the same order of typical 
height fluctuations over a distance $L_k$.
In the RG procedure, we shall fix $h_k=2W(L_k)$ in order
to have a scale invariant description of the surface
(see ref. \cite{CML} for details).
The coarse-graining procedure, which defines the static 
description in terms of blocks of size $L_k$, also induces 
a flow of the microscopic dynamics towards an {\em effective 
dynamics} at the same scale $L_k$; this is defined in terms 
of the transition rates for the addition of an occupied block.
The main feature of KPZ dynamics is lateral 
growth\cite{kpz,mmtb}, and this suggests the following 
minimal parametrization of the growth rates at the 
generic scale $L_k$ is
\be
r[h(i)\to h(i)\!+\! 1] \equiv 1\!+\!x_k \!
\sum_{j nn i}\max[0,h(j)-h(i)].
\label{rates}
\ee
The first term is the contribution of the vertical growth (i.e. 
random deposition) and  the second term is the contribution of 
lateral growth. Indeed the sum over neighbor block sites $j$ 
counts the area of the vertical surface exposed towards site $i$.
$x_k$ is then the ratio of lateral to vertical growth rates.
We shall come back later, in the conclusions, to the approximations
implied by Eq.~(\ref{rates}).

In order to derive the RG transformation, let us 
partition a system of size $L$ into $\ell^d$ cells
of size $L/\ell$. We observe that the roughness $W^2(L)$ can
be written as the average roughness  $W^2(L/\ell)$ {\em inside} single cells
plus the fluctuations of the average height {\em among} different cells.
The second contribution, within the description at scale $L/\ell$, 
is simply given by the roughness $\omega^2(\ell,x)$ of a system 
of $\ell$ cells - with $x$ being the dynamic parameter at scale
$L/\ell$ - times the square height of the cells $[2W(L/\ell)]^2$. 
Hence we find
\be
W^2(L)=W^2(L/\ell)\left[1+4\omega^2(\ell,x)\right]
\label{W2}
\ee
which is the basis of the RG approach.
With $\ell=4$, $L=L_{k+2}$ and $x=x_k$, it gives 
$W^2(L_{k+2})/W^2(L_k)$. The same quantity can be alternatively
computed using Eq.~(\ref{W2}) with $\ell=2$, once
with $L=L_{k+2}$ and $x=x_{k+1}$, and a second time with 
$L=L_{k+1}$ and $x=x_k$. The consistency of the two 
calculations yields an implicit RG transformation
\be
1+4\omega^2(2,x_{k+1})=\frac{1+4\omega^2(4,x_k)}
{1+4\omega^2(2,x_k)}
\label{f4}
\ee
for the dynamic parameter $x_k$. 
The {\em attractive} fixed point $x^*=\lim_{k\to\infty} x_k$ 
(if it exists) identifies the {\em scale invariant dynamics}
and
Eq.~(\ref{W2}) with $x_k=x^*$ finally yields the roughness exponent
\be
\alpha = \lim_{k \rightarrow \infty} \log_2  \sqrt{
\frac{W^2(L_{k+1})}{W^2(L_{k})} } =  
\frac{\log [1+4\omega^2(2,x^*)]}{2\log 2}.
\label{f5}
\ee
Eqs. (\ref{f4}, \ref{f5}) are the starting point of our
analysis. A more detailed discussion of their derivation 
can be found in Ref. \cite{CML}. We note here that the 
existence of an attractive fixed point $x^*$ implies that 
the process is ``self-organized'': No fine tuning 
is necessary in order to observe the critical behavior. 

A key observation is that, since $\omega^2(\ell,x)\to 0$ 
for $x\to\infty$, $x^* = \infty$ is a fixed point of 
Eq.~(\ref{f4}) corresponding to $\alpha=0$.
Therefore the RG scheme allows, in 
principle, for the occurrence of a finite upper critical
dimension $d_c$ ($\alpha=0$ for $d\ge d_c$) and
the existence of a finite attractive fixed point for all $d$ is 
a non-trivial prediction.
A finite stable fixed point was found in Ref. \cite{CML} 
for $d=1,\ldots,8$ using Monte Carlo methods to compute 
$\omega^2(\ell,x)$. The same method was also applied to the 
Gaussian theory [$\lambda=0$ in Eq.~(1)], recovering
the result $d_c=2$, i.e. $\alpha=0$ for $d\ge 2$
\cite{unpub}. Though very powerful, the Monte Carlo method
cannot be pushed to very high dimensions nor does it 
provide an explicit analytic behavior of $\alpha$ as a 
function of $d$. 

In the following we study analytically the large-$d$ limit 
of the RG in order to extract its predictions on the 
existence of a finite upper critical dimension and on the
large-$d$ behavior of the roughness exponent. 

The technical difficulty lies in the explicit calculation
of the functions $\omega^2(\ell,x)$ for $\ell=2,~4$. 
For $d\gg 1$ we expect $\alpha\ll 1$, which means that
surface fluctuations $\omega(\ell,x)\sim \ell^\alpha\simeq 1+\alpha
\ln\ell+\ldots$ are of order $1$. 
This suggests that for a system of small size $\ell$
we can reasonably account for the fluctuations of the interface
if we allow $h(i)$ to take only two values: $h(i)=h_0$ or 
$h(i)=h_0+1$. This drastic approximation has the advantages
of making the explicit computation feasible on the one hand, and
of providing a lower bound for the exponent $\alpha$ on the
other. We shall come back later to this important issue.
Let us only stress, for the time being, that a lower
bound on $\alpha$ is sufficient to exclude the existence of 
a finite $d_c$.

In the above approximation, growth can only occur 
on ``low'' sites ($h(i)=h_0$). This means that Eq.~(\ref{rates}) 
is only valid if $h(i)=h_0$ and the rates vanish on ``high'' sites
($h(i)=h_0+1$). It is convenient to classify the possible configurations
$\{h(i)\}$ by the number $n$ of ``high'' sites.
The roughness Eq.~(\ref{roughness})
of each configuration of $n$ ``high'' sites is the same and is equal to 
$(1-n/\ell^d)n/\ell^d$ and the dynamics involves only transitions 
from configurations with $n$ to configurations with $n+1$ ``high''
sites.
We can then group all configurations$\{h(i)\}$ with $n$ ``high''
sites in the same effective state with a great simplification of
the structure of the master equation
(the state with $n=\ell^d$ is equivalent to the flat surface $n=0$).
The only non-vanishing transition rates $r(n\to n+1)$ are
obtained from Eq.~(\ref{rates}) summing on all possible final
configurations and taking the average on the initial configurations,
which leads to
\be
r(n \rightarrow n +1)= \ell^d-n+x\,\Omega_n.
\label{rates1}
\ee
The first term here accounts for vertical growth, which can occur only
on the $\ell^d-n$ ``low'' sites.
The second term is the contribution of lateral growth and
$\Omega_n$ is the average
number of lateral walls (i.e. the surface between ``low'' and
``high'' sites) in configurations with $n$ ``high'' sites. 
Assuming that ``low'' and ``high'' sites are randomly
distributed, each ``low'' site has on average 
$2d n/\ell^d$ ``high'' neighbor sites and therefore
\be 
\Omega_n \simeq 2d (\ell^d -n) \frac{n}{\ell^d}.
\label{random}
\ee
The distribution of ``high'' sites is actually not random
but we have verified numerically that, for large enough dimensions,
Eq.~(\ref{random}) provides a reasonable 
approximation\cite{nota}.
Combining Eqs. (\ref{rates1}, \ref{random}) one easily 
obtains the probability $\rho_n$ of state $n$ in the stationary state
of the master equation
\be
\rho_n = \rho_{0} \frac{r(0 \rightarrow 1)} {r(n \rightarrow n+ 1)},
~~~~n=1,\ldots,\ell^d-1,
\label{ro}
\ee
where $\rho_0$ is fixed by the normalization condition
$\sum_{n=0}^{\ell^d-1} \rho_n =1$. A simple
calculation leads to
\be 
\rho_0 = \left\{  1 + \frac{\ell^d}{2 d x_k} \left[ 
2 d \ln \ell + \ln  \left(\frac{1+2dx_k}{\ell^d+2 d x_k} 
\right) \right]
          \right\}^{-1}.
\label{ro0}
\ee
The roughness of configurations with $n$ particles, 
using Eq.~(\ref{roughness}), is $(1-n/\ell^d)n/\ell^d$ which,
averaged over the distribution $\rho_n$ (as specified by Eq.~(\ref{ro})
and (\ref{ro0})) gives
\be 
\omega^2(\ell,x)\cong \rho_0\frac{\ell^d}{2dx_k}
\label{f6}
\ee
where we have assumed $dx_k \gg 1$ and $\ell^d\gg 1$.
Combining Eq.~(\ref{f6}) with the RG equation (\ref{f4})
we obtain, to leading order in $d$, a fixed point
\be
x^* =  2^{d+1}\ln 2.
\label{xstar}
\ee
consistent with the assumption $dx_k \gg 1$.
Using now Eq.~(\ref{f5}) it is straightforward
to find, to leading order in $d\gg 1$,
\be
\alpha \simeq  \frac{1}{3(\ln 2)^2}\frac{1}{d}.
\label{alpha}
\ee
Furthermore, we can also analyze the stability of the 
fixed point. The derivative of the RG transformation
$x_{k+1}=R(x_k)$ of Eq.~(\ref{f4}), at the fixed point, is
\be
R'(x^*)=-1+\frac{1}{2\ln 2}\frac{1}{d}+O(d^{-2})
\ee
Since $|R'(x^*)|<1$ we can conclude that the fixed point 
is attractive $\forall d$. 
Therefore {\em we find a finite, stable fixed point $x^*$ with 
an exponent $\alpha>0$ for all $d$}, which is the 
main result of this Letter. This excludes the occurrence
of a finite upper critical dimension, $d_c$, which would 
show up, in the present framework, in a stable fixed
point at $x^*=\infty$ for $d\ge d_c$.

Let us now discuss the validity of the approximations used.
We neglected configurations with $h(i)\ge h_0+2$ or equivalently
deposition processes on a ``high'' site. 
The rate of this process, on a state
with $n$ ``high'' sites, is $r_{\rm up}(n)= n$. Our approximation 
is then valid if $r(n\to n+1)\gg r_{\rm up}(n)$. This condition
fails when the process is close to complete a new layer, i.e.
for $n\simeq \ell^d$. More precisely the deposition on ``high''
sites is not important for 
\be
\ell^d-n \ge 1 \gg \frac{\ell^d}{2dx}.
\label{condition}
\ee
Since $x^*\sim 2^d$, the approximation is correct
for $\ell=2$ $\forall d$.
Fig. \ref{fig2} shows the that the approximation to
the LHS of Eq.~(\ref{f4}) $1+4\omega^2(\ell=2,x)$ is 
good already for $d=7$.
The approximation is much less accurate for $\ell=4$ and,
as a consequence, fluctuations in the system of size
$\ell=4$ are underestimated. This means that our 
approach underestimates the RHS of Eq.~(\ref{f4}) and 
consequently also its value at the intersection point
with the LHS. This value is directly related to the 
roughness exponent by Eq.~(\ref{f5}) and therefore
{\em the restriction of height fluctuations leads to a
lower bound to the exponent $\alpha$}.
Accordingly - since the LHS of Eq.~(\ref{f4}) decreases with $x$ -
Eq.~(\ref{xstar}) gives an upper bound to
the true fixed point parameter $x^*$. Fig. \ref{fig2}
illustrates this analysis for $d=7$. 
Fig. \ref{fig3} shows a comparison of the present analytical
estimates [Eqs. (\ref{xstar}) and (\ref{alpha})] and the
results of ref. \cite{debate}, \cite{Ala} and \cite{CML}.

Besides the approximations of the present calculation, which,
as we have argued, provide a lower bound to $\alpha$, it is
also worth discussing the approximations of the RG method
itself. In this respect we observe that Eq.~({\ref{rates}) 
is a minimal parametrization of the dynamics, in the sense 
that it allows for the minimal proliferation in the
RG capturing the relevant features of KPZ growth. 
In principle, more proliferation parameters can be included in
order to improve the accuracy of the method. It is important to
note, however, that the range of typical fluctuations 
$h(i)-h(j)\sim \ell^\alpha$ allowed in the RG calculation 
is small and the one-parameter approximation in Eq.~(\ref{rates}) 
to the scale invariant dynamics is reasonable. This is 
confirmed by the accuracy of the RG predictions in finite dimensions 
\cite{CML} and it is expected to improve as $\alpha\to 0$.
Therefore the inclusion of additional proliferation parameters in
Eq.~(\ref{rates}) is not expected to change the nature of the fixed 
point and of our main conclusions. Let us also point out that usually 
small cells analysis becomes very accurate in high dimensions.
An extension of the RG procedure to cells of larger size,
going beyond the present approximations, provides in principle
a systematic way to improve our prediction which is
currently under investigation~\cite{unpub}.

In conclusion we have shown that the recently proposed\cite{CML}
real space RG predicts that the roughness exponent $\alpha$
decreases not faster than $1/d$ as $d\to\infty$ [Eq.~(\ref{alpha})]. 
This implies that there is no finite upper critical dimension 
in the KPZ universality class and it suggests that theoretical 
arguments leading to $d_c=4$ should be reconsidered. 


We acknowledge interesting discussions with A. Maritan, G. Parisi,
A. Stella, C. Tebaldi and A. Vespignani.

\begin{figure}
\centerline{\psfig{figure=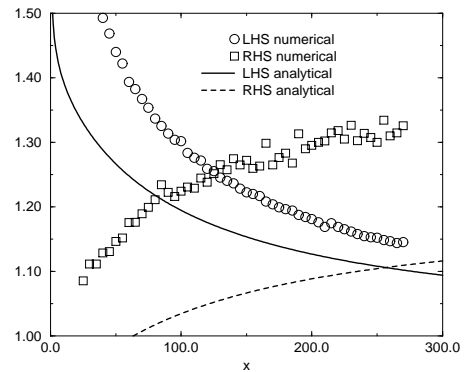,width=7cm}}
\caption{
Graphic analysis of Eq.~(5) from the present approximation and 
from Monte Carlo evaluation for $d=7$.
}
\label{fig2}
\end{figure}
                                                      
\begin{figure}
\centerline{\psfig{figure=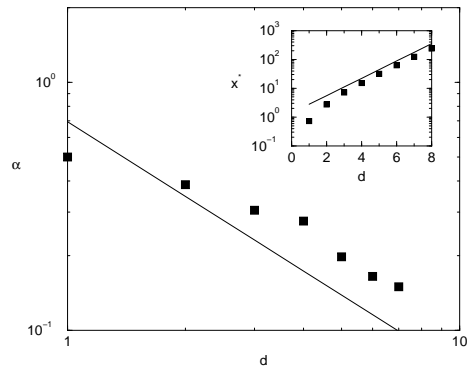,width=7cm}}
\caption{ Value of $\alpha$ from the present calculation (full line)
and from simulations of ref.~8 and 18 for $d=1,\ldots,7$. In view
of the approximations involved (see text) we obtain a lower bound.
Inset: fixed point value $x^*(d)$ vs $d$. The theoretical prediction
(full line) is an upper bound to the true $x^*(d)$.
}
\label{fig3}
\end{figure}
                                            

\begin{thebibliography}{99}
\bibitem{nota}
``High'' sites rather than being randomly distributed, tend 
to coalesce in droplets. This means that the approximation 
is worst for $n\sim \ell^d/2$. This effect is not very important
for small sizes $\ell$ in $d\gg 1$.
\end{thebibliography}
\end{document}